\documentclass[a4paper,12pt]{article}
\usepackage{amssymb}
\usepackage{amsmath}
\usepackage{latexsym}
\usepackage{epsfig}
\usepackage{graphicx}
\usepackage{amsmath,amsthm,amsfonts,amssymb}
\usepackage{subfig}

\newcommand{\bi}{\bibitem}
\newcommand{\be}{\begin{eqnarray}}
\newcommand{\ee}{\end{eqnarray}}

\newcommand{\BH}    {{\rm BH}}
\newcommand{\weak}  {{\rm W}}

\newcommand{\CP}    {{\rm CP}}

\newcommand{\SM}    {{\rm SM}}

\newcommand{\DW}    {{\rm DW}}
\newcommand{\sph}   {{\rm sph}}

\begin{document}
	
	\begin{titlepage}
		%\begin{flushright}
		%hep-th/yymmnnn\\ August
		%\end{flushright}
		\begin{centering}
			\vspace{.8in}
			{\large {\bf Electroweak baryogenesis by primordial black holes in Brans-Dicke modified gravity}}
			\\
			
			\vspace{.5in}
			{\bf Georgios Aliferis\footnote{aliferis@physics.auth.gr},
				\bf Vasilios Zarikas\footnote{vzarikas@uth.gr}}$^{,3}$

			\vspace{0.3in}
			$^{1}$ Department of Physics\\
			Aristotle University of Thessaloniki, 54124 Thessaloniki, Greece\\
			\vspace{0.3cm}
			$^{2}$ 
			General Dept. University of Thessaly, 35100 Lamia, Greece\\
			\vspace{0.3cm}
			$^{3}$ School of Engineering\\
			Nazarbayev University, Nursultan, Republic of Kazakhstan, 010000\\
		\end{centering}
		
		\vspace{1in}

		%\author{Georgios Aliferis}\email{aliferis@physics.auth.gr}
		%\affiliation{Department of Physics,\\ Aristotle University of Thessaloniki,
		%54124, Thessaloniki, Greece}

		%\author{Vasilios Zarikas}\email{vzarikas@teilam.gr}
		%\affiliation{Department of Electrical Engineering, Theory Division\\ ATEI Lamias,
		%35100 Lamia, Greece}

		\begin{abstract}
			
			A successful baryogenesis mechanism is proposed in the cosmological framework of Brans-Dicke modified gravity. Primordial black holes with small mass are produced at the end of the Brans-Dicke field domination era. The Hawking radiation reheats a spherical region around every black hole to a high temperature and the electroweak symmetry is restored there. A domain wall is formed separating the region with the symmetric vacuum from the asymmetric region where electroweak baryogenesis takes place. First order phase transition is not needed. In Brans-Dicke cosmologies, black hole accretion can be strong enough to result to cosmic black hole domination, extension of the lifetime of black holes and enhanced baryogenesis. The analysis of the whole scenario, provides very easily and without fine tuning the observed baryon number asymmetry for either small or big CP-violating angles in the finite temperature corrected effective potential of Two-Higgs Doublet Models. The advantage of our proposed scenario with Brans-Dicke modified gravity is that naturally provides both black hole domination and efficient baryogenesis for smaller CP violating angles compared to the same mechanism applied in a FRW cosmological background.  
			
		\end{abstract}
		
	\end{titlepage}
	
	\newpage
	
	\baselineskip=18pt
	
\section{Introduction}

An important still open issue for cosmology is baryogenesis. For baryon number to be produced, three criteria must be satisfied, as stated by Sakharov \cite{sakharov}:\\
1. Baryon number non-conservation.\\
2. C and CP symmetry violation.\\
3. Out of thermal equilibrium conditions.

Many baryogenesis models have been produced over the last decades (reviews \cite{review1} - \cite{review6}). The majority of these studies work with a baryon number violation that occurs either at grand unification \cite{gut-B} or at the electroweak energy scale \cite{ew-B}.

One scenario (\cite{sakharov}, \cite{kuzmin} - \cite{ad-yz}) is the baryon asymmetry to be produced by heavy particles decay in an expanding universe, with C and CP symmetry broken. These heavy particles can be gauge bosons of a grand unified theory. A problem with these models is that the baryon number produced can be wiped out in some later process, as sphaleron processes at $\sim$ 100GeV.

Electroweak baryogenesis is another possibility \cite{krs}, \cite{CKN}. Chiral anomaly is a cause for baryon number violation \cite{ew-B}. The phase transition of the electroweak breaking could be of first or second order. However, in the Standard Model (SM), the transition proved to be second order; the large value of Higgs mass killed any hopes for first order transition and thus the net baryon number produced if any, is destroyed by sphalerons. Another problem for standard model electroweak baryogenesis is that it predicts CP-violating angles smaller than required \cite{ad-varenna}. The electroweak baryogenesis can also be combined with some modified gravity theory, like TeV scale gravity \cite{TeV-gravity}, \cite{TeV-grav-EW}.

Baryo-through-leptogenesis \cite{fuku-yana} refers to lepton number production by heavy Majorana particles decay, at energies high as $10^{10}$GeV. The lepton asymmetry then leads to baryon asymmetry through electroweak processes that violate the (B+L) symmetry \cite{lgns}. Some other possibilities are Affleck - Dine \cite{affleck} and spontaneous \cite{spont-BS}  baryogenesis.

Baryon asymmetry can also be produced by primordial black holes (PBH) \cite{zeld-bh}. PBHs could be created at the beginning of the universe \cite{carr}, \cite{Khlopov}. Initially, it was considered that PBHs can generate baryon excess by GUT processes, \cite{GUT}. The problem with this, as with other GUT baryon number violating models, is that the baryon asymmetry created can be washed out later by sphaleron processes \cite{sphaleron}, as we have explained. An interesting model of electroweak baryogenesis by PBHs was proposed by Nagatani \cite{Nagatani}. According to this, the baryon excess is produced in a thermal domain wall that separates a reheated, by Hawking radiation, area around the PBHs from the outer regions, where $T<100GeV$. Other, also worth mentioning models, which incorporate electroweak baryogenesis around PBHs have been proposed \cite{EW_PBH_baryog}. Electroweak baryogenesis by PBHs becomes very efficient \cite{AZK} in the case of high energy modifications of Hubble rate in the early universe, as in Randall - Sundrum cosmology \cite{RS}. 

In the present paper, we propose a novel model of electroweak baryogenesis by PBHs in Brans-Dicke (BD) cosmology. We assume that the early Universe starts from either primordial black hole dominated era or from an initially radiation dominated era with mixture of radiation and primordial black holes. Brans-Dicke theories can realize such a scenario. While universe temperature has been lowered below electroweak symmetry breaking point ($\sim 100GeV$), a region around each PBH is reheated by Hawking radiation to $T\,>\,100GeV$. A domain wall is formed between the symmetric and asymmetric regions and this is where baryogenesis takes place, by sphaleron processes. The key characteristics of the baryogenesis scenario are:\\
1. The EWK phase transition can be of second order and the non equilibrium conditions are due to the formation of the domain wall around PBHs. The baryon over anti-baryon excess is created by sphalerons.\\
2. In order to produce the observed baryon number with small CP violating angles ($b/s\simeq 6\times10^{-10}$), the universe needs to become PBH dominated. In BD - cosmology this may happen naturally, because of accretion by the PBHs. In standard cosmology, on the contrary, it is accepted that accretion may not be so strong as in modified gravities \cite{zeldo}.\\
3. The CP-violating angle must be larger compared with the one in SM for adequate baryogenesis. This can be satisfied incorporating any phenomenologically viable two Higgs doublet with CP phase at hight temperatures, instead of a single Higgs model.
	
Brans-Dicke gravity, \cite{BD}, or otherwise named Jordan Brans Dicke, \cite{J}, is a modified gravity theory \cite{Clifton:2011jh}. Its difference from general relativity (GR) is that the gravitational constant G is not constant. Instead, its value is the inverse of a time-dependent scalar field $\phi$. This $\phi$ couples to gravity with a coupling parameter $\omega$. When $\omega \to \infty$ BD becomes GR. Solar system measurements require $\omega \gtrapprox 10^{4}$. In conventional BD, $\omega$ is constant and so this present time limit holds also for the very early universe. Nevertheless, there are generalizations of the BD theory where $\omega$ varies with time, \cite{BDW}. Its present value may obey the above limit, but may be much smaller during the early universe. Another class of generalised BD theories is that of the complete BD theories, \cite{CBD}. They incorporate energy exchange between the scalar field and ordinary matter. In the present study we work for simplicity, to be able to derive semi-analytical results, with the conventional BD gravity. However, since our baryogenesis happens in the early cosmic history we present results allowing the free parameter $\omega$ to take values both smaller and larger than $10^{4}$. So we assume that after baryogenesis the cosmic evolution is determined better by another BD model that relaxes the constraints on $\omega$ in the very early Universe.

PBHs are created at the end of the BD - field ($\phi$) domination era; however, the model is not dependent on how they were created. Accretion can lead to  BHs mass increase only when there is enough radiation for BHs to accrete. This may happen during an initially radiation dominated era or even during an initial BH - domination time period, if there is enough radiation density, as we are going to show. Thus, two cases are examined: the first is that the universe becomes BH dominated immediately after PBHs creation, with BH density $\rho_{BH}=0.7\rho$ and radiation density $\rho_{rad}=0.3\rho$. The second is the case that PBHs are initially, immediately after their formation, only a small part of the universe but then, because of intense accretion, become dominant. It will be shown that for both cases there is a range of initial PBHs masses for which accretion leads the universe to become completely BH dominated ($\rho_{BH}\simeq 100\%$).

The advantage of the proposed scenario is that  Brans-Dicke gravity, due to enhanced accretion, can naturally provide black holes domination in the early Universe and at the same time, as we are going to show, efficient baryogenesis for smaller CP-violating angles compared to the case of the same scenario but with the gravity of General Relativity. 

In the following section, the baryon asymmetry mechanism is described. In section 3 we analyse the fist of the two cases of the proposed scenario, a black hole dominated Universe, while in section 4 we study a Universe that initially is radiation dominated but then becomes black hole dominated. Next a section with various bounds is given. A study of non trivial mass spectrum is also analysed and finally the last section provides a conclusive summary. 

\section{Baryon number created by a single primordial black hole}

The PBHs of our proposed mechanism are surrounded by radiation colder than the electroweak breaking point ($T_W \sim 100GeV$). They are very small and thus Hawking temperature $T_{BH}$ is much greater than this temperature. Then all kinds of Standard Model particles are emitted and they are in symmetric phase. So, the Hawking emission causes the thermalization of the black hole surrounding region.
A local temperature $T(r)$ can be defined for a region with size greater than the mean free path (MFP) of the emitted particles. The MFP of a particle $f$ is $\lambda_f(T) = \frac{\beta_f}{T}$, where $\beta_f$ is a constant that depends on the particle species. Quarks and gluons have a strong interaction and they have the shortest MFP with $\beta_s \simeq 10$. Because of the high, larger than $T_{EW}$, reheating temperature, all SM particles contribute to the massless degrees of freedom ($g_{*\SM} \equiv \sum_f g_{*f} = 106.75$). So, the radiation density is $\rho=\frac{\pi^2}{30} g_{*\SM} T^4(r)$. In this section we follow the analysis resented in \cite{Nagatani} and references therein. Some improvements of this analysis are also presented and are clearly pointed out.  

The closest outer region to the PBH horizon, with length up to the MFP of the quarks and gluons, is not thermalized. For this reason, the emitted particles move freely there and most of them don't drop back to the black hole. Thus, the black hole radiation obeys the law of Stefan - Boltzmann with not significant corrections. Now, let $r_{o}$ be the minimum thermalized radius and $T_o$ the local temperature there: $T_{o}=\frac{\beta_s}{r_o}$. We consider then the transfer equation of the energy in the thermalized region to determine the temperature distribution $T(r)$. We assume diffusion approximation of photon transfer at the deep light-depth region is valid \cite{Mihalas}. The diffusion current of energy in Local Temperature Equilibrium (LTE) is $J_\mu = - \frac{\beta}{3\;T(r)} \: \partial_\mu \rho$. The quantity $\beta/T$ is the effective MFP of all particles by all interactions with $\beta \simeq 100$. The transfer equation is $\frac{\partial}{\partial t} \rho = - \nabla_\mu J^\mu$. A stationary spherical-symmetric solution \cite{Mihalas} is
\begin{equation}
T(r)^3 =  T_{bg}^3 + \frac{r_{o}}{r}\,(T_{o}^3 - T_{bg}^3)  \, .
\end{equation}
where $T_{bg}$ is the background temperature. It can be as high as somewhat lower than $T_{EW}$, where sphaleron rate is suppressed.

The quantities $r_{o}$ and $T_{o}$ can be written as functions of black holes temperature $T_{BH}$ by equalizing the outgoing diffusion flux $4\pi r^2 J(r)\simeq \frac{8\pi^3}{135} \beta_s \beta \,{g_{*\SM}}\;[1 - (T_{bg}/T_{o})^3]\;T_{o}^2$ with the Hawking radiation flux $4\pi r_\BH^2 \times \frac{\pi^2}{120} {g_{*\SM}}T_\BH^4$:
\begin{eqnarray}
r_{o} &=& \frac {16 \pi} {3}\,\frac{1}{T_\BH} \sqrt{  \beta_{s}^3\beta \,  [1 - (T_{bg}/T_{o})^3] }
\end{eqnarray}
and
\begin{eqnarray}
T_o &=& \frac{3}{16\pi\sqrt{\beta_s \beta }} \;\frac{T_\BH}{\sqrt{1 - (T_{bg}/T_o)^3}} \, .
\end{eqnarray}
$T_{bg} \ll T_o$ and so the spherical thermal distribution surrounding the black hole for $r > r_{o}$ is
\begin{eqnarray}
T(r)^3 &=& 
T_{bg}^3 + \frac{9}{256\pi^2}\frac{1}{\beta}\frac{T_\BH^2}{r}.
\label{ghw}
\end{eqnarray}

%%%%%%%%%%%%%%%%%%%%%%%%%
%% Domain Wall
%%%%%%%%%%%%%%%%%%%%%%%%%

As mentioned before, the region around PBHs is reheated to temperatures higher than the electroweak breaking point and so symmetry is restored there. The background temperature, at the same time, remains below the electroweak breaking point and the symmetry broken. That means that an electroweak domain wall forms around the black hole and it starts at $r_{DW}$. The phase transition at the domain wall does not have to be of first order. It can be a second order transition. This enlarges the parameter space of the validity of our proposed scenario.  

%%%TWO HIGGS%%%%

Instead of a single Higgs SU(2) doublet, we incorporate a two Higgs doublet model (2HDM) in our proposed mechanism, since it can accommodate a CP-violation in the Higgs sector. The present study does not depend on a specific 2HDM. The only requirement is the existence of a CP violating phase in the finite temperature corrected effective potential of the Higgs sector. As an example, such a model of 2HDM that could fit in our scenario is the work, \cite{2HDM2017}. This is a concrete, phenomenologically correct model that can also provide large thermal corrected CP violating angles. However, it is worth mentioning at this point that it will be shown that our mechanism provides efficient baryogenesis for both small and large CP violating phases. Of course, if one 2HDM provides larger CP angles this is more than welcome since it enlarges the allowed set of free parameters for efficient baryogenesis.

The tree-level, CP-breaking scalar potential in \cite{2HDM2017} is
\begin{eqnarray}
V_{tree} &=& m _{11}^{2}\Phi _{1}^{\dagger}\Phi_{1}+ m_{22}^{2}\Phi _{2}^{\dagger }\Phi _{2} - \left[m_{12}^2 \Phi_{1}^{\dagger}\Phi_{2} +h.c.\right] +\frac{1}{2} \lambda_{1}(\Phi _{1}^{\dagger }\Phi_{1})^{2} +\frac{1}{2}\lambda_{2}(\Phi _{2}^{\dagger }\Phi _{2})^{2}   \nonumber
\\
&& +\lambda_{3}(\Phi_{1}^{\dagger}\Phi_{1}) (\Phi_{2}^{\dagger}\Phi_{2}) +\lambda_{4}(\Phi_{1}^{\dagger}\Phi_{2}) (\Phi_{2}^{\dagger}\Phi_{1}) +\left[{\frac{1}{2}}\lambda_{5}(\Phi_{1}^{\dagger}\Phi_{2})^{2} +h.c.\right]\, , \label{pote}
\end{eqnarray}
where
\begin{equation}
\Phi _{1}=\left( {\
	{{\phi_{1}^+ \atop \phi_{1}^0}}}\right) ,
\;\;\Phi _{2}=
\left( {\ {{\phi_{2}^+ \atop \phi_{2}^0}}}\right)
\label{fi}
\end{equation}
are the two $SU(2)_L$ scalar field doublets. One can see that a $Z_2$ discrete symmetry holds, under which $\Phi_{1}\rightarrow \Phi_{1}$ and $\Phi_{2}\rightarrow -\Phi_{2}$. Because of this symmetry there are no flavour changing neutral currents. The symmetry is softly broken only by $m_{12}$. The parameters of the potential are real, because of its hermiticity, except from the mass parameter $m_{12}$ and the quartic coupling $\lambda_5$. With this scalar potential it is possible the doublets VEVs to be complex and this CP-violation cannot be gauged away due to the complex values of $m_{12}$ and $\lambda_5$.

The proposed baryogenesis scenario we study does not depend on a specific 2HDM model. We just need any type of Two-Higgs model (real or CP violating 2HDM) with CP violation at high temperatures in the phase of one of the two Higgs doublets fields at high temperatures. The reason we mention this model, \cite{2HDM2017}, is that this is a concrete recent minimal model; it presents the finite temperature corrections in a clear way and at hight temperatures can provide big CP violating angle of order one, $O(1)$, without any phenomenological problems.  
The present scenario can give efficient baryogenesis for both small and large CP violating angles. Thus, our proposed scenario can fit with any 2HDM with CP violating phase in any of the two doublets at hight temperatures. However, big CP violating angles are always welcome since in this way the set of allowed parameter space is enlarged. 

Note that in our previous work \cite{AZK} a different 2HDM had been adopted. In that work we had additional D-breaking terms $ V_{{\rm D}}=\lambda _{6}(\Phi
_{1}^{\dagger }\Phi _{1})(\Phi _{1}^{\dagger }\Phi _{2})+\lambda _{7}(\Phi
_{2}^{\dagger }\Phi _{2})(\Phi _{1}^{\dagger }\Phi _{2})\,+\,h.c.
\label{soft}
$, with parameters
$\lambda _{6}$ and $\lambda _{7}$, in general complex numbers.

We can simplify the form of the doublets with an $SU(2)$ rotation. $\partial V/\partial \phi _{i}=0$ solutions give stationary points, including the asymmetric minimum that respects the $U(1)$ of electromagnetism:
$
\Phi _{1}={\frac{1}{\sqrt{2}}}\left( {\
	{{0 , u}}}\right)^{\intercal} ,
\;\;\Phi _{2}={\frac{1}{\sqrt{2}}}
\left( {\ {{0 , ve^{i\varphi }}}}\right)^{\intercal} .
$
where $u,v,\varphi$ are real and $\varphi$ is the CP-violating angle. This tree-level CP-violating phase depends on $m_{12}$ and $\lambda_5$ and cannot be shifted by an SU(2) rotation or with another allowed physical gauge. 

In the review paper of \cite{branco}, one can see the three different types of minima in 2HSM like the one we need. You cannot gauge away the phases that appear in these minima for the tree potential that we are using with independent values of the imaginary part of $m_{12}$ and $\lambda_5$. However, in this case, we need this CP angle to be very small due to Electron Dipole Moment constraints (EDM) \cite{edm}. To  achieve strong CP-violations one can hope the loop finite temperature corrected potential to result to big CP-violating cases \cite{loop},\cite{Zarikas:1999bf, Zarikas:1995qb, Lahanas:1998wf}. In this case, the constraints from EDM do not apply if at zero temperature the CP angle goes to very small values. The possible mimima/saddle points at tree level are of course related with the minima or saddle points that appear in the loop finite temperature potential, \cite{Zarikas:1995qb}, \cite{2HDM2017} since they are the temeperature evolution of them.  

Regarding the cosmological consequences, anyway, the finite temperature effective potential is this that should be used. The temperature loop corrections incorporate for the larger range of the parameters space only small cubic resulting to a second order phase transition (in \cite{2HDM2017} the case of first order transition is also studied, something that is not needed in our scenario). We shift the scalar fields about their expectation values and the second doublet asymmetric minimum becomes

\begin{equation}
\Phi _{1}={\frac{1}{\sqrt{2}}}\left( {\
	{{0 \atop u(T)}}}\right) ,
\;\;\Phi _{2}={\frac{1}{\sqrt{2}}}
\left( {\ {{0 \atop v(T)\,e^{i\varphi(T) }}}}\right) .
\label{I}
\end{equation}
with
\begin{equation}
v(T) = v \;f(r)\; , \label{VEV.eq}
\end{equation}
where $f(r)$ is a form-function of the wall and has a value from zero to one; $f(r)=0$ for $r \leq r_\DW$ and $f(r)=\sqrt{1 - \left(\frac{T(r)}{T_\weak}\right)^2}$ for  $r > r_\DW$.

At the limit between the thermalized sphere and the domain wall, the temperature is  $T(r_\DW) = T_{\weak}$. Setting this in Eq. (\ref{ghw}), we find the radius of the thermalized region $r_\DW$. The width of the domain wall $d_\DW$ is about of the order of $r_\DW$.
\begin{equation}
d_\DW \simeq r_\DW
= \frac{9}{256 \pi^2\,\beta_{br}} \frac{1}{1 - (T_{bg}/T_\weak)^3}\frac{T_\BH^2}{T_\weak^3} \, .
\end{equation}

The structure of the electroweak domain wall is determined only by the thermal structure of the black hole and not by the dynamics of the phase transition as in the ordinary electroweak baryogenesis scenario (the CKN model).

%%%%%%%%%%%%%%%%%%%%%%%%%%%%%%%%%%%%%%%%%%%%%
\section{First case: Black hole domination from the moment of creation}
%%%%%%%%%%%%%%%%%%%%%%%%%%%%%%%%%%%%%%%%%%%%%

 In our model, the universe at the beginning of its life is dominated by the BD - field. We assume that the PBHs creation happens at about the end of this period. Then the universe becomes a mixture of radiation and black holes. The free parameters of our model are: Number of black holes, $N$, initial value of black hole density, $\rho_{BH}(t_i)$, initial value of time $t_i$, the initial black hole mass  $m_{BH}(t_i)=m_i$ and $\omega$, the characteristic parameter of Brans-Dicke gravity. The characterization "initial", means after full black hole creation at the end of the BD domination era. Theoretically it could be possible to relate the number of black holes, $N$, and initial value of black hole density, $\rho_{BH}(t_i)$, through a relation like $\rho_{BH}(t_i)=N\,m_i\,H(t_i)^3$. Then one can specify the initial Hubble rate through the current one value via the cosmic time evolution. However, practically this cannot be done since in our proposed scenario (i) we "believe" and follow the BD gravity only in the early cosmic evolution and we assume another modified gravity, BD like gravity, with varying $\omega$ to be true theory for lower energies (ii) there are ambiguities related with the value of the parameter $teq$, $t_1$ (see below).  
 
 A first scenario we examine is that universe is BH dominated immediately after $t_i$, that is $\rho_{BH}\,>\,\rho_{rad}$. It becomes completely BH dominated because of accretion, if their initial masses are above the mass limit that accretion exceeds evaporation and the radiation is dense enough, as it will be shown. What follows is that having no more radiation to accrete, they only evaporate. The quantity, $t_{ev}$, is the time of complete evaporation. The universe then turns radiation dominated, with the observed baryon number already produced. Later the universe turns from radiation to dust dominated at $t_{eq}$. It remains dust dominated until now ($t_0$).

Barrow and Carr at \cite{BarrowCarr} have obtained solutions for G for the three different eras of a model where the universe is initially dominated by the BD - field, then it turns radiation dominated and finally dust dominated:

\begin{eqnarray}
G(t) &\simeq& G_0 (\frac{t_1}{t})^{\sqrt{n}}(\frac{t_0}{t_{eq}})^{n}, \,\, if\,\,t<t_1: BD - field \,\, dominated \nonumber\\
&& G_0 (\frac{t_0}{t_{eq}})^{n},  \,\, if\,\, t_1<t<t_{eq}: radiation \,\, dominated \nonumber\\
&& G_0 (\frac{t_0}{t})^{n}, \,\, if\,\, t_{eq}<t: dust \,\, dominated
\end{eqnarray}
where $t_1$ is the time of transition from BD - field dominated to radiation and $n=\frac{2}{4+3\omega}$. To avoid confusion it is worth mentioning that there is no PBHs - domination era at Barrow - Carr work.

The modified solutions for our model are:
\begin{eqnarray}
G(t) &\simeq& G_0 (\frac{t_i}{t})^{\sqrt{n}}(\frac{t_0\,t_{ev}}{t_{eq}\,t_i})^{n}, \,\, if\,\, t<t_i: BD - field \,\, dominated \nonumber\\
&& G_0 (\frac{t_0\,t_{ev}}{t\,t_{eq}})^{n}, \,\, if\,\, t_i<t<t_{ev}: PBHs \,\, dominated \nonumber\\
&& G_0 (\frac{t_0}{t_{eq}})^{n}, \,\, if\,\, t_{ev}<t<t_{eq}: radiation \,\, dominated \nonumber\\
&& G_0 (\frac{t_0}{t})^{n}, \,\, if\,\, t_{eq}<t: dust \,\, dominated
\label{G}
\end{eqnarray}
Then we need to write formulas for universe density due to PBHs ($\rho_{BH}$) and scale factor $\alpha$. The number density of PBHs at the time of their creation is:
\begin{equation}
n_{BH}(t_i)=\frac{\rho_{BH}(t_i)}{m_{BH}(t_i)}
\end{equation}
BHs can be treated as dust, regarding the universe 's density due to them. Because of the fact that their mass changes due to accretion and evaporation, it is their number density $n_{BH}(t)$, not density, that is inversely proportional to scale factor 3rd power, and so:
\begin{equation}
\rho_{BH}(t) = n_{BH}(t) m_{BH}(t) =
 n_{BH}(t_i)\frac{\alpha^3 (t_i)}{\alpha^3 (t)} m_{BH}(t) = \rho_{BH}(t_i) \frac{m_{BH}(t)}{m_{BH}(t_i)} \frac{\alpha^3 (t_i)}{\alpha^3 (t)}
 \label{robh }
\end{equation}
We assume that the number of black holes after their primordial creation and till their evaporation remains the same. Thus, we assume that these PBHs do not "eat" each other in a considerable rate during the accretion. Accretion concerns the surrounding radiation mainly.

In \cite{BarrowCarr} can be found the scale factor's time evolution:
\begin{eqnarray}
\alpha(t) &\propto& t^{(1-\sqrt{n})/3}, \,\, BD - field \,\, dominated \nonumber\\
&& t^{1/2}, \,\, radiation \,\, dominated \nonumber\\
&& t^{(2-n)/3}, \,\, dust \,\, dominated
\label{a}
\end{eqnarray}
and so

\begin{equation}
\rho_{BH}(t) =\rho_{BH}(t_i) \frac{m_{BH}(t)}{m_{BH}(t_i)} (\frac{t_i}{t})^{2-n}
\label{robhdom}
\end{equation}
The radiation density at the same time will be $\rho_{rad}(t) = \rho(t) - \rho_{BH}(t)$.

Now we can have a formula for $\rho(t)$ solving the first Friedmann equation. Friedmann equations for $k=0$ (flat universe) and including the BD - field $\phi$ are:
\begin{equation}
\frac{\dot{\alpha}^2}{\alpha^2}+\frac{\dot{\alpha}}{\alpha}\frac{\dot{\phi}}{\phi}-\frac{\omega}{6}\frac{\dot{\phi}^2}{\phi^2}=\frac{8\pi\rho}{3\phi} \nonumber\\
\end{equation}
\begin{equation}
2\frac{\ddot{\alpha}}{\alpha}+\frac{\dot{\alpha}^2}{\alpha^2}+2\frac{\dot{\alpha}}{\alpha}\frac{\dot{\phi}}{\phi}+\frac{\omega}{2}\frac{\dot{\phi}^2}{\phi^2}+\frac{\ddot{\phi}}{\phi}=-\frac{8\pi p}{\phi} \nonumber\\
\end{equation}
\begin{equation}
\frac{\ddot{\phi}}{8\pi}+3\frac{\dot{\alpha}}{\alpha}\frac{\dot{\phi}}{8\pi}=\frac{\rho-3p}{2\omega+3}
\label{Fried}
\end{equation}
Then we can use Eq. (\ref{a}) for dust domination:
\begin{equation}
\frac{\dot{\alpha}}{\alpha} = \frac{2-n}{3}\, t^{-1} \label{dotatoa}
\end{equation}
and also Eq. (\ref{G}) for BH domination:
\begin{equation}
\phi = \frac{1}{G(t)} = \frac{1}{G_0} (\frac{t_{eq}}{t_0\, t_{ev}})^{n}\, t^{n} \Rightarrow \frac{\dot{\phi}}{\phi} = n\, t^{-1} \label{dotphitophi}
\end{equation}
Substituting these and also $\omega = \frac{2}{3n} - \frac{4}{3}$ to the first Friedmann equation, it becomes:
\begin{equation}
\rho(t) = \frac{n+4}{24 \pi G_0} (\frac{t_{eq}}{t_0\, t_{ev}})^n \,t^{n-2}
\end{equation}

To calculate the baryon number produced by each one PBH, we have to know how their mass evolves with time due to accretion and evaporation.
\begin{equation}
\dot{m}_{acc} = 4 \pi f R_{BH}^2 \rho_{rad}
\label{acc}
\end{equation}
where f is accretion efficiency of order $O(1)$. We set $f=2/3$, as in \cite{BDMaju}. $R_{BH}=2Gm_{BH}$ is the radius of the BH.
\begin{equation}
\dot{m}_{ev}= -4 \pi R_{BH}^2 a_H T_{BH}^4 = -\frac{a_H}{256 \pi^3} \frac{1}{G^2 m_{BH}^2}
\label{ev}
\end{equation}
where $a_H$ is an effective Stefan-Boltzmann constant. It is defined as $a_H = \frac{\pi^2}{120}g_{*SM}$, where we remind that $g_{*SM}$ is the massless degrees of freedom considering all Standard Model particles massless. This is so because BH temperature is higher than the EW scale.
Combining accretion and evaporation and using G(t) from Eq. (\ref{G}) for BH-domination, we get:

\begin{equation}
\dot{m}_{BH} = 16 \pi f G_0^2 (\frac{t_0\, t_{ev}}{t_{eq}})^{2n} t^{-2n} m_{BH}^2 \rho_{rad} -
\frac{a_H}{256 \pi^3} \frac{1}{G_0^2} (\frac{t_0\, t_{ev}}{t_{eq}})^{-2n} t^{2n} m_{BH}^{-2}
\label{Mrate}
\end{equation}

At this point in order to analyze the whole scenario, we have to set some indicative values to our free parameters. Since we want to study a black hole dominated Universe from the moment of PBHs domination we select $\rho_{BH}=2\,\rho_{rad}$.  We proceed with calculations for $\omega=10^4$, which is the observational limit for the present value of $\omega$. In another section we will also present results for different values of $\omega$, since this is meaningful as we have explained in the introduction.

\begin{figure}
	% Requires \usepackage{graphicx}
	\includegraphics[scale=1]{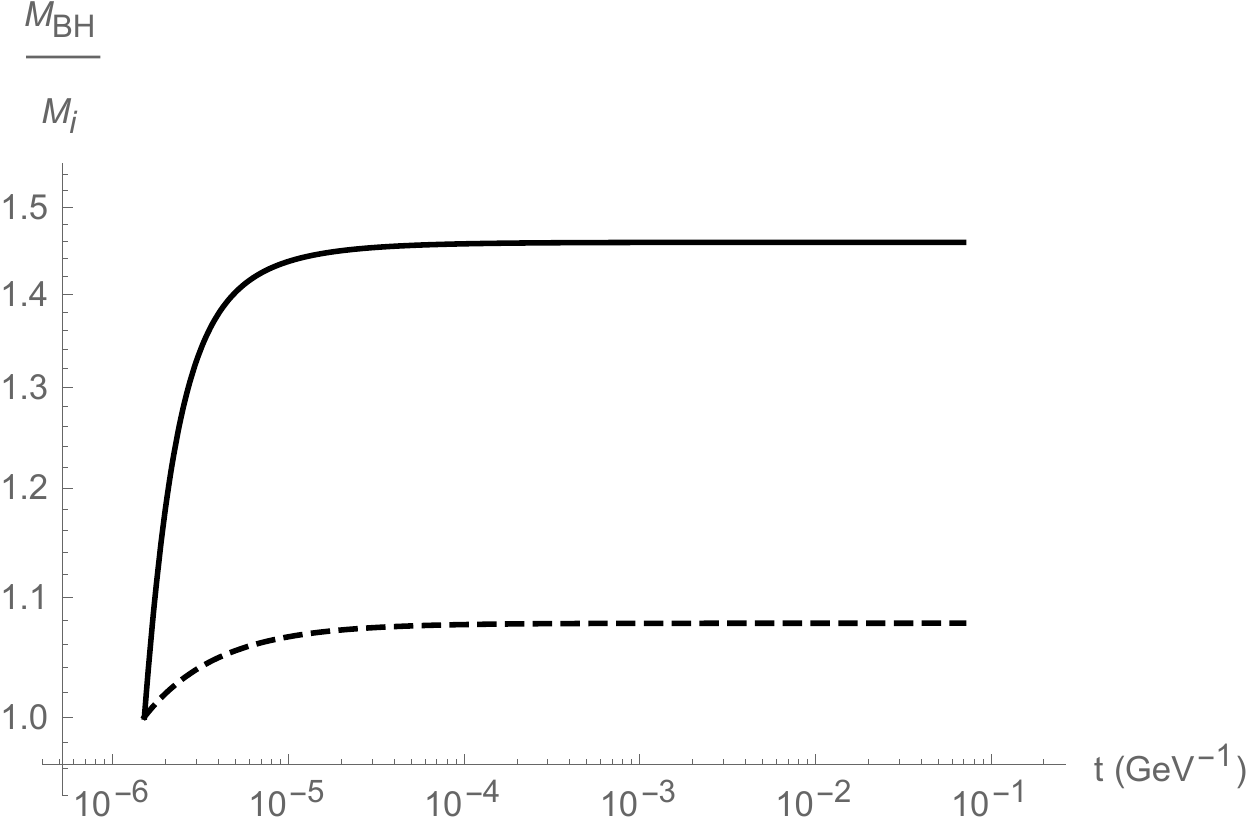} 
	\caption{$t_i = 10^{-30} sec,\,\rho_{BH}(t_i) = 0.67\rho(t_i)$: accretion is not efficient for initial masses up to $10^{31}GeV$ (dashed line). It is sufficient to lead to almost total BH domination for $10^{32} GeV$ (continuous line). } \label{7030-303231}
\end{figure}

In order to have a feeling about the black hole masses that are relevant for our scenario we demand $\dot{m}_{BH}=0$ and we find the initial BH mass for which  accretion equals evaporation,  $m_i\, \simeq\, 10^{25} GeV$ (or $\,\simeq \,1gr$) for $t_i\,=\,10^{-30}sec$. Yet, for initial PBH masses up to  $m_i\,=\,10^{31}GeV$, as shown in Fig. (\ref{7030-303231}), accretion is able to increase the mass of the PBH only a little at the beginning. This is so because the radiation that was to be eaten becomes rapidly less dense, due to the universe expansion. Only $10^{32}GeV$ or greater values lead BHs to accumulate almost the entire universe mass (Fig. \ref{7030-303231})). Note that  the value $m_{BH}=10^{32}GeV$ is an upper limit, as it will be shown in the bounds section.

\begin{figure} %
	\centering
	\subfloat[\centering accretion]{{\includegraphics[width=7cm]{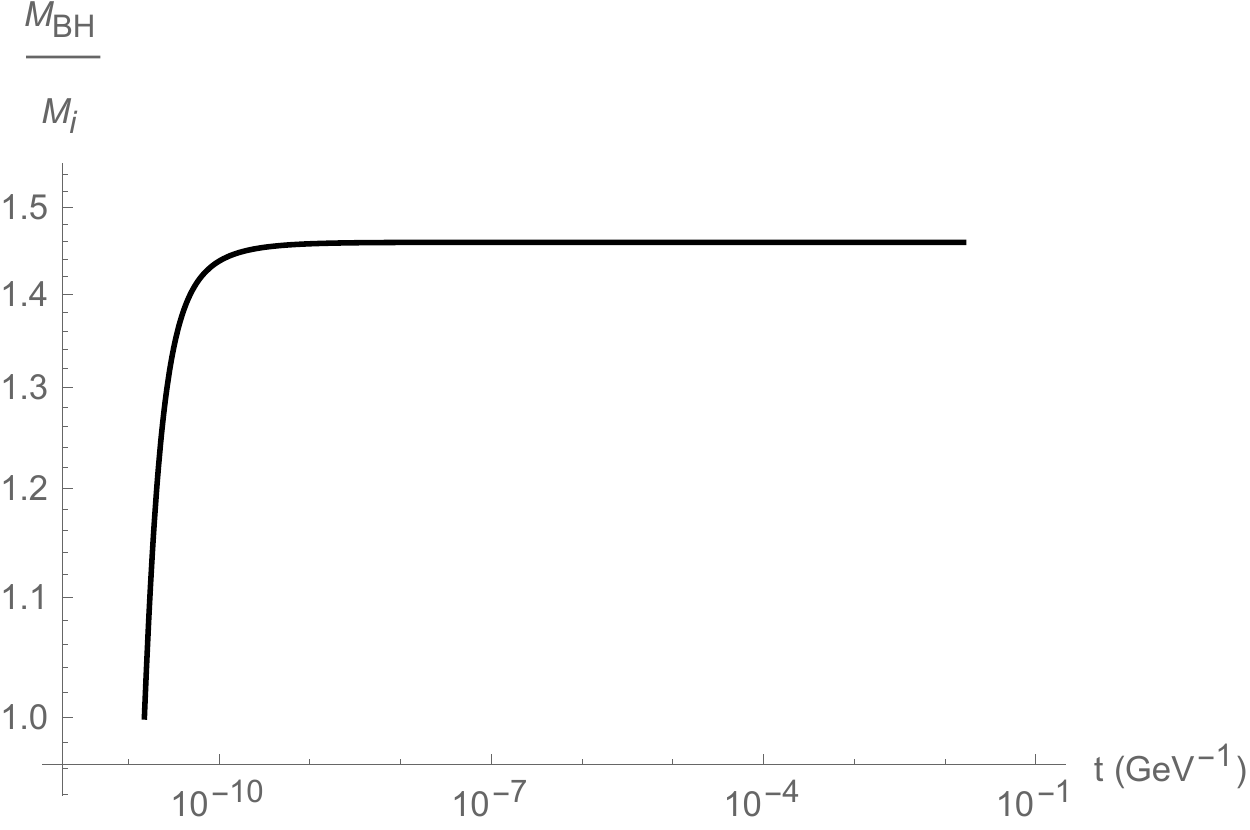} }} %
	\qquad
	\subfloat[\centering evaporation]{{\includegraphics[width=7cm]{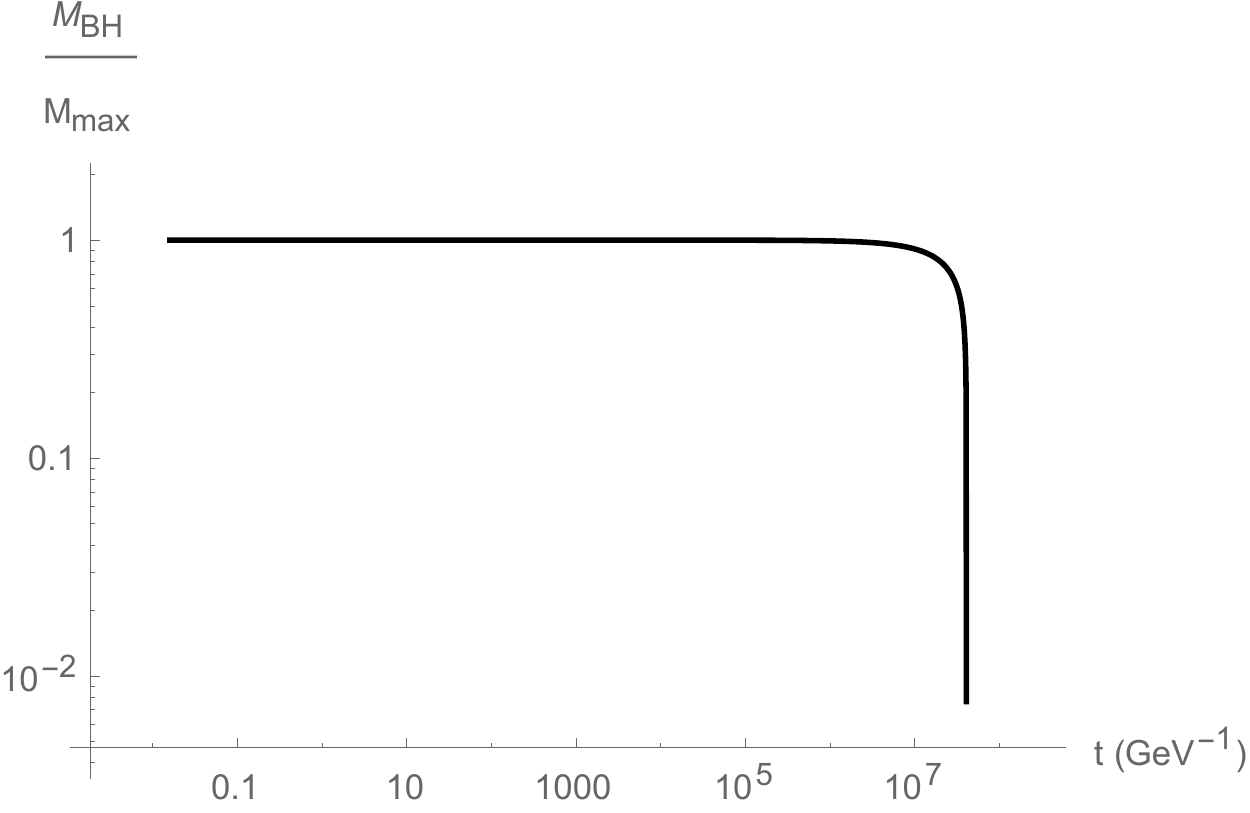} }}%
	\caption{$t_i = 10^{-35} sec,\,\rho_{BH}(t_i) = 10^{-3}\rho(t_i)$. For $m_i=10^{27} GeV$ or greater, accretion is sufficient to lead to almost total BH domination.}%
	\label{7030-3527} %
\end{figure}

Things are different in the case that PBH creation takes place earlier: $t_i\,\simeq\,10^{-35}sec$. Initial accretion now equals evaporation for $m_i=2.7\times10^{22}GeV$. Denser radiation makes accretion strong enough to lead to almost complete ($98\%$) BH domination, for smaller initial masses ($m_i\geq 10^{27}GeV$, Fig. (\ref{7030-3527})). Then there is no more radiation for accretion to proceed.

The time that evaporation becomes stronger than accretion is given from Eq. (\ref{Mrate}), for $\dot{m}_{BH}=0$ and for $m_{BH}= m_{max}$. For the $m_i = 10^{27} GeV ,\, t_i = 10^{-35} sec$ case it is $m_{max} \sim 1.45 \times 10^{27} GeV$ and $t_{acc=evap} \sim 10^{-26}sec$. Universe will turn to radiation dominated with the evaporation of the PBHs. The evaporation and the result for the values in regard is shown in Fig. (\ref{7030-3527}).
The time of complete evaporation is for $m(t)=0$ and it is $t_{ev}\simeq 2.7 \times 10^{-17}sec$.

%%%%%%%%%%%%%%%%%%%%%%%%
\subsection{Baryogenesis}
%%%%%%%%%%%%%%%%%%%%%%%%

In the following, we calculate the baryon number generated by a single black hole and then the baryon to entropy ratio $b/s$ of the universe.

Although sphaleron process takes place both in the symmetric region around a black hole and the domain wall, the required CP - violation and non-equilibrium conditions coexist only in the domain wall. So, it is there that the baryon assymetry is created. In addition, $f(r)=|\langle\phi_{2}(r)\rangle|/v \leq \epsilon =1/100$ is needed, so as the order of the sphaleron process exponential factor to be one and the baryon asymmetry not to be suppressed. In other words,  baryon generation happens in the region of the domain wall that Higgs scalar value is small and this is from $r_{DW}$ to $r_{DW}+d_{sph}$, where $d_{sph}$ is defined from $f(r_{DW}+d_{sph})=\epsilon$. Then, it is $ \int_{r_\DW}^{r_\DW+d_\sph} dr \: \frac{d}{dr} \varphi(r)=\epsilon\,\Delta\varphi_\CP $, where
$\varphi(r,T)=[f(r)-1]\Delta\varphi_\CP $, \cite{CKN}. Thus,
\begin{eqnarray}
\dot{B}
&=&  V \; \frac{\Gamma_\sph}{T_\weak} \; {\cal N} \dot{\varphi}
\nonumber\\
&=&  4\pi {\cal N} \kappa\: \alpha_\weak^5 T_\weak^3 \;
r_\DW^2 \; v_\DW
\int_{r_\DW}^{r_\DW+d_\sph} dr \: \frac{d}{dr} \varphi(r)
\nonumber\\
&=&  \frac{1}{16\pi} \: {\cal N} \kappa\: \alpha_\weak^5 \:
\epsilon\:\Delta\varphi_\CP \:
\frac{T_\BH^2}{T_\weak}
\label{BRate}
\end{eqnarray}
where $\Gamma_{sph}$ is the sphaleron transition rate, $\Delta\varphi_\CP$ the net CP phase.  ${\cal N} \simeq O\left(1\right)$ is a model dependent
constant which is determined by the type of spontaneous electroweak baryogenesis scenario and the fermion content,
$\kappa \simeq O\left(30\right)$ is a numerical constant expressing the strength of the sphaleron process

Integrating numerically through the BHs lifetime, we calculate the total baryon number by a single BH.
\begin{equation}
B = \int_{t_i}^{t_{ev}}  \dot{B} \: dt
\label{BR2}
\end{equation}
The baryon number produced during accretion is orders of magnitude smaller than during evaporation.

After BHs have gained their maximum mass, they only evaporate at a slow rate until the last moments before their complete annihilation (Fig. (\ref{7030-3527})). Thus, in \cite{AZK} we used (and also Y. Nagatani in \cite{Nagatani}. However note that there was also an error regarding the black hole density in \cite{Nagatani} as we explain in \cite{AZK}) an approximation where BHs mass remains constant until the time of evaporation when it turns to radiation completely. In this approximation the total baryon number density produced was evaluated by:
\begin{equation}
b = B \, \frac{\rho_{BH}(t_{ev}^{-})}{m_{max}}
\label{b}
\end{equation}
where
\begin{equation}
\rho_{BH}(t_{ev}^{-}) = \rho_{rad} \left(t_{ev}^{+} \right) = \frac{\pi^2}{30}g_{reh}\,T_{reh}^4 \,,
\label{rhorho}
\end{equation}
and $T_{reh}$ is the temperature that the universe is reheated as BHs evaporate (with a typical choice of $T_{reh} = 95 GeV$ so as to be below $T_W$). Eq. (\ref{rhorho}) is approximately true since the last part of the evaporation happens very quickly and we suppose that the radiation that has not been eaten by accretion is negligible.
However, the expression Eq. (\ref{b}) is problematic.
There is an ambiguity with what black hole mass to divide in Eq. (\ref{b}). The mass in the denominator can take values from $m_{max}$ till zero when the evaporation completes and the baryon asymmetry takes is larger value. During the rapid evaporation the mass decreases from the maximum value to zero and the same happens for the black hole number density. For this reason we propose an advanced estimation:
\begin{equation}
b=B\, N\, H^3(t_{ev}) \, ,
\end{equation}
where $N$ is the number of black holes and $H^{-3}(t_{ev})$ a measure of the volume of the universe at the time of complete evaporation, that is the reheating. Since the baryogenesis is completed at the end of evaporation, at that moment, the total baryon asymmetry that have been produced should be diluted with this volume.
We can calculate $H(t^-_{ev})$ from the BD cosmology expansion.

Now, the entropy density is, \cite{Kolb}, 
\begin{equation}
s = \frac{2\pi^2}{45}g_{reh}T_{reh}^3\,\, 
\end{equation}
where $g_{reh}$ is the massless degrees of freedom of the reheated plasma in the asymmetric phase.

Requiring at least the observed $b/s \sim 6 \times 10^{-10}$ we calculate the value of the free parameter N in terms of the CP-violating phase $\Delta \theta_{CP}$:

\noindent For the case $t_i=10^{-30}s$, $m_i=10^{32}GeV$ it is $N\Delta \theta_{CP} \geq 5.8\times10^{50} $.

\noindent For the case $t_i=10^{-35}s$, $m_i=10^{27}GeV$ it is $N\Delta \theta_{CP} \geq 5.8\times10^{10} $.

\noindent So, it turns out that we can have the observed $b/s$ even for small values of $\Delta \theta_{CP}$, since $N$ is a free parameter. The constraints on N are discussed in the bounds section.

%%%%%%%%%%%%%%%%%%%%%%%%%%%%%%%%%%%%%%%%%%%%%%%%%%%%%%%%%%
\section{Second case: Primordial black holes domination because of accretion}
%%%%%%%%%%%%%%%%%%%%%%%%%%%%%%%%%%%%%%%%%%%%%%%%%%%%%%%%%%

Another case, even more interesting, is the one where the PBHs, at the end of their creation, are only a small fraction of the total universe density and the universe is radiation dominated. As it will be shown, accretion can be strong enough to lead to PBH domination and the production of the observed baryon number.

Thus, in this scenario, PBHs the end of the $\phi$-domination era, consist only a portion of $\rho$. That means a radiation domination period begins after the $\phi$-domination era. If accretion is strong a PBH domination epoch follows, after $t_{eq1}$. Time, $t_{eq1}$, is the moment BHs density becomes equal to radiation density. The universe turns radiation dominated for the second time after BHs evaporation.

Therefore, the evolution of $G(t)$ now is (if accretion lead from radiation to PBH domination):
\begin{eqnarray}
G(t) &\simeq& G_0 (\frac{t_i}{t})^{\sqrt{n}}(\frac{t_0\, t_{ev}}{t_{eq}\, t_{eq1}})^{n}, \,\, if\,\, t<t_i: BD - field \,\, dominated \nonumber\\
&& G_0 (\frac{t_0\, t_{ev}}{t_{eq1}\, t_{eq}})^{n}, \,\, if\,\, t_i<t<t_{eq1}: radiation \,\, dominated \nonumber\\
&& G_0 (\frac{t_0 \,t_{ev}}{t\, t_{eq}})^{n}, \,\, if\,\, t_{eq1}<t<t_{ev}: PBHs \,\, dominated \nonumber\\
&& G_0 (\frac{t_0}{t_{eq}})^{n}, \,\,  if\,\,t_{ev}<t<t_{eq}: radiation \,\, dominated \nonumber\\
&& G_0 (\frac{t_0}{t})^{n}, \,\,  if\,\,t_{eq}<t: dust \,\, dominated
\label{G2}
\end{eqnarray}
For the period $t_i<t<t_{eq1}$, $G$ is constant, as one can see from Eq. (\ref{G2}), and so $\dot{\phi}=0$. Then, the first Friedmann equation (Eq. (\ref{Fried})) becomes:
\begin{equation}
\frac{\dot{\alpha}^2}{\alpha^2}=\frac{8\pi\rho}{3\phi}
\end{equation}
where $\alpha\propto t^{1/2}$ and so we can evaluate the time evolution of the total density $\rho(t)$. Furthermore, Eq. (\ref{robh }) for the PBHs energy density holds. Substituting the corresponding $a$:
\begin{eqnarray}
\rho_{BH}(t) &=&\rho_{BH}(t_i) \frac{m_{BH}(t)}{m_{BH}(t_i)} (\frac{t_i}{t})^{3/2},\,\,\,t_i<t<t_{eq1} \nonumber\\
&=&\rho_{BH}(t_{eq1}) \frac{m_{BH}(t)}{m_{BH}(t_{eq1})} (\frac{t_{eq1}}{t})^{2-n},\,\,\,t_{eq1}<t<t_{ev}
\end{eqnarray}
For the radiation part it is still  $\rho_{rad}(t) = \rho(t) - \rho_{BH}(t)$.

\begin{figure} %
	\centering
	\subfloat[\centering accretion]{{\includegraphics[width=7cm]{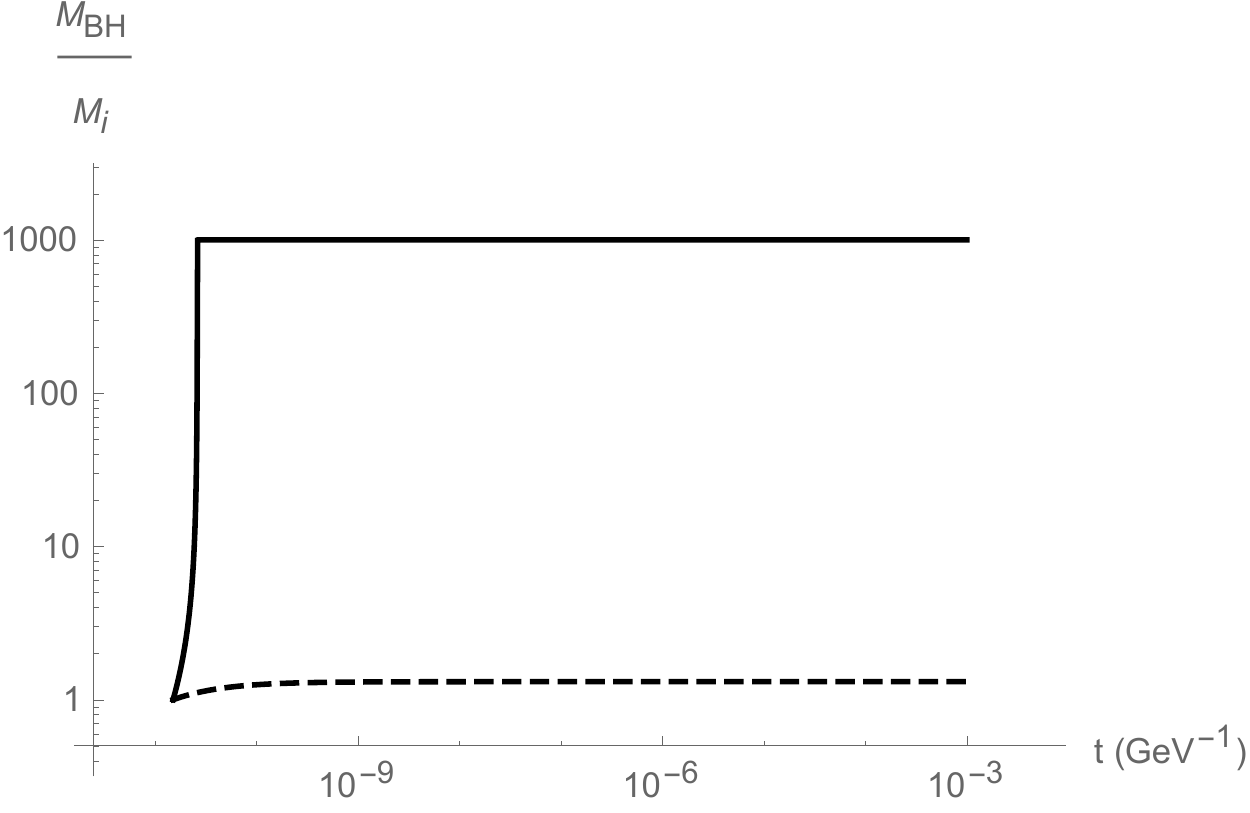} }}%
	\qquad
	\subfloat[\centering evaporation]{{\includegraphics[width=7cm]{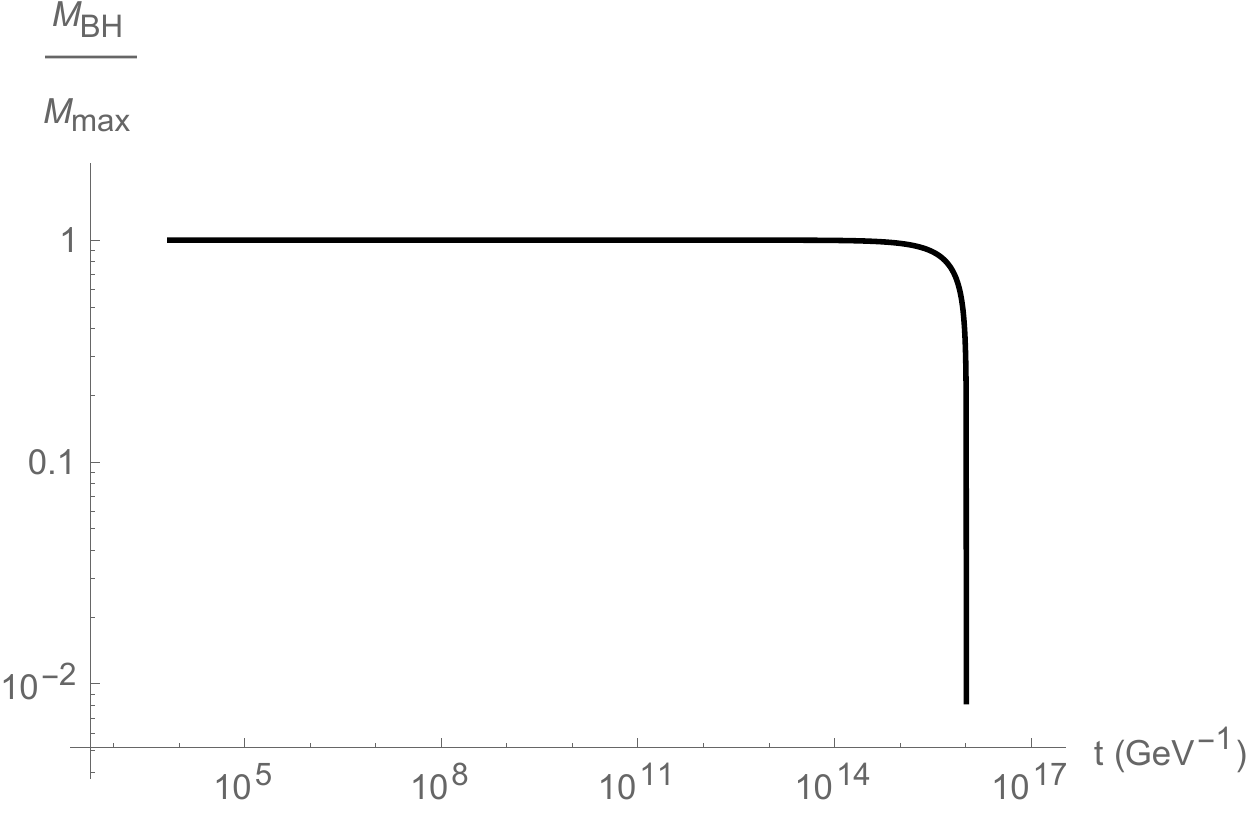} }}%
	\caption{$t_i = 10^{-35} sec,\,\rho_{BH}(t_i) = 10^{-3}\rho(t_i)$: accretion is not efficient for initial masses up to $10^{26}GeV$ (dashed line). It is sufficient to lead to almost total BH domination for masses equal to $10^{27}GeV$ (continuous line) or greater.}%
	\label{1999-352726}%
\end{figure}

The mass evolution of the PBHs is determined, as in the previous case, by accretion, Eq. (\ref{acc}), and evaporation, Eq. (\ref{ev}). We proceed with calculations for $\omega=10^4$, which is the observational limit for the present value of $\omega$. More results for different values of $\omega$ will follow. The limit for accretion to be stronger than evaporation is now $m_{BH}(t_i) \sim 2 \times 10^{22} GeV$. In Fig. (\ref{1999-352726}) is shown the mass evolution during the accretion period. One can see that accretion is very effective for $m_i \geq 10^{27} GeV$ and as a consequence the universe becomes almost completely PBH dominated. With $95\%$ of the density inside the BHs , there is nothing else to accrete. Evaporation follows, see Fig. (\ref{1999-352726}).

 So, the mass of a single PBH can increase up to 100,000 times, from $10^{27}$ to $10^{32}$ GeV (which is an upper bound), because of accretion. The black hole lifetime also increases because of the mass increase.
 
 The mechanism of baryon number production is the same as in the previous case and thus the baryonic asymmetry created by a single PBH is considerably enhanced. The total baryon number to entropy density is calculated as a function of the free parameter $N$ and the CP violation angle $\Delta \theta_{CP}$. For the case we examined of the minimum initial mass $10^{27}GeV$ that leads to total BH domination ($t_i=10^{-35}s$, $m_i=10^{27}GeV$, $\rho_{BH}(t_i)=10^{-3}\rho(t_i)$), it is $N\Delta \theta_{CP}\geq2.4\times10^{33} $ for the observed $b/s \geq 6\times 10^{-10}$. Again, the observed $b/s$ can be obtained even for small values of $\Delta \theta_{CP}$.
 
 %%%%%%%%%%%%%%%%%%%%%%%%%%%%%%%%%
 \section{$\omega$-dependence}
 %%%%%%%%%%%%%%%%%%%%%%%%%%%%%%%%%
 
 It is interesting to examine now how the value of the $\omega$ affects the baryogenetic mechanism. It is anticipated that the lower the value the easier the accretion by the black holes. We remind that the limit Brans-Dicke gravity meets General Relativity is for $\omega \to \infty$, while observations impose $\omega \gtrsim 10^{4}$ for the present time.
 
 In the investigation carried in the previous sections we found that the initial PBH mass leading to total PBH domination through accretion is $m_i \gtrsim 10^{27} GeV$ for $t_i=10^{-35}s$ and $\omega = 10^{4}$. For $\omega = 10^{10}$ now, a high value that makes BD-gravity almost identical to GR, and all the other parameters the same, accretion is inefficient. The BH mass increases only a little. It turns out that $m_i \gtrsim 10^{28} GeV$, for accretion leading to full PBH domination. So, accretion is more efficient in BD-gravity than in GR.
 
 Moreover, as we have mentioned, the observational limit for the present value of $\omega$ does not have to hold for the past if we assume that a BD gravity with time varying $\omega$ is the correct theory. Since the proposed mechanism concerns only a very short time duration of the very early universe, we are going to examine even very low $\omega$ values using the formalism of the conventional BD gravity. 
 
 At Table (\ref{ttable1}) we consider some characteristic cases where PBHs are born at $t_i=10^{-35}s$ and their initial masses are the lowest that lead to complete BH domination. We have taken $\rho_{BH}(t_i)=10^{-3} \rho(t_i)$ for all the cases in order the results to be more easily comparable. We made an exception for $\omega=1$ ($\rho_{BH}(t_i)=10^{-15} \rho(t_i)$) because the maximum masses of the PBHs after the accretion have to be inside the range $10^{28} GeV < m_{BH} < 10^{32} GeV$ (see next section).
 
 \begin{table}
 \begin{center}
 	\begin{tabular}{cccc}
 		$\omega$ & $m_i\,(GeV)$ & $m_{max}\,(GeV)$& $ N\times\Delta\theta_{CP}$ \\\hline
 		$10^{10}$ & $10^{28}$ & $10^{31}$& $6.2\times 10^{35}$ \\
 		$10^{4}$ & $10^{27}$ & $10^{30}$& $2.4\times 10^{33}$ \\
 		$10$ & $10^{26}$ & $10^{29}$& $1.1\times 10^{22}$ \\
 		$1$ & $10^{17}$ & $10^{32}$& $2.5\times 10^{52}$ \\
 	\end{tabular}
 \end{center} 
\caption{Lower values of $\omega$ lead to enhanced accretion and thus lower initial PBH masses result to total BH domination.}
\label{ttable1}
\end{table}

It is apparent that the lower the $\omega$ the more efficient the accretion as it leads to complete PBH domination for lower initial PBH masses. Especially for values close to $1$ it becomes extremely efficient. It can drive PBHs with initial mass as low as $10^{17}GeV$ to increase by a factor of up to $10^{15}$ (this is the case at the last line of Table (\ref{ttable1})).

The dependence of accretion on $\omega$ is similar in the case of PBH-domination from the beginning. In Table (\ref{ttable2}) we show the maximum increase of the PBH mass for a characteristic case ($t_i=10^{-30}s$, $\rho_{BH}(t_i)=0.67 \rho(t_i)$, $m_i=10^{31}GeV$) where the maximum possible mass enhancement is $1.5$, when PBHs consume all the energy of the universe.

\begin{table}
\begin{center}
	\begin{tabular}
		{cc}
		$\omega$ & $m_{max}/m_i$ \\\hline
		$10^{10}$ & $1.04$ \\
		$10^{4}$ & $1.08$ \\
		$10$ & $1.25$ \\
		$1$ & $1.5$ \\
	\end{tabular}
\end{center}
\caption{$t_i=10^{-30}s$, $\rho_{BH}(t_i)=0.67 \rho(t_i)$, $m_i=10^{31}GeV$: accretion is enhanced as $\omega$ decreases.}
\label{ttable2}
\end{table}

 %%%%%%%%%%%%%%%%%%%%%%%%%%%%%%%
 \section{Bounds}
 %%%%%%%%%%%%%%%%%%%%%%%%%%%%%%%
 
 One limit for PBHs mass is posed by the fact that the size of the domain wall $d_{DW}$ must be greater than the mean free path (MFP), $\lambda=\beta_S /T_W$. Since 
 \begin{equation}
 d_{DW} = \frac{9}{256 \pi^2}\frac{1}{\beta_{SM} c_W} \frac{T_{BH}^2}{T_W^3}
 \end{equation}
 it is needed
 \begin{equation}
 d_{DW} > \lambda \Rightarrow T_{BH} > 53 TeV \Rightarrow m_{BH} < 10^{32} GeV.
 \end{equation}

 Another limit appears because the black hole lifetime $\tau_{BH}$ should be quite greater than the time for the stable weak domain wall to form. The evaporation equation (Eq. (\ref{ev})) is integrated analytically:
 \begin{equation}
 \label{evap}
 m(t) = \frac{3^{\frac{1}{3}}((f-1)a_H\, t_0^{-2n}\, t_{eq}^{2n}\, t_{ev}^{-2n}\, t^{1+2n}+ 256\, G_0^2\, \pi^3 (1+2n)m_{max}/3)^{1/3}}{4 \times 2^{2/3}\, G_0^{2/3}\, (1+2n)^{1/3}\pi}
 \end{equation}
 The formula for BH lifetime without accretion was used (that is from $m_{max}$ till complete evaporation) because the time period from the moment that evaporation starts to dominate is orders of magnitude greater than the time length of the dominant accretion period. At this point we should point out that for all interesting parameter space the accretion happens very rapidly and we rich the maximum black hole mass; after there is long period where evaporation is dominant (accretion at some point ends because there is not radiation left to be eaten) but the black hole mass decreases very slowly till a very sudden rapid annihilation of all the black hole mass. 
 \begin{equation}\label{evaptime}
 \tau_{BH} \sim \frac{256 \, G_0^2 \, m_{max}^3\, (1+2n)\, \pi^3\, t_0^{2n}\, t_{eq}^{-2n}}{3\,a_H (1-f)} \, .
 \end{equation}
 The domain wall formation time is
 \begin{equation}
 \tau_{DW} = \frac{d_{DW}}{u_{DW}} = \frac{27\, T_{BH}^4}{4096\, \pi^4\, \beta_{SM}^3\, c_W^3\, T_W^5}
 \end{equation}
 Solving for $m_{max}$ we find that it should be, approximately $ \tau_{BH}>\tau_{DW}\Rightarrow m_{max}\, >\, 10^{28} \,GeV\,$. The masses that provide successful baryogenesis in our model are within these limits. To avoid confusion, this second constraint provides a lower bound on masses. The parameter $m_{max}$ refers to the maximum value after accretion finishes to be dominant. 
 
 One more constraint but this time for the BH density, can be obtained demanding the Universe after complete BHs evaporation to reheat at least to nucleosynthesis temperature. 
 After the black holes evaporation the universe is reheated, its density is in the form of radiation and equals $\rho_{\rm rad}(t_{\rm reh})=\rho_{\rm reh}=\frac{\pi^2}{30} g_{\rm reh} T^{4}_{\rm reh} $,
 roughly the minimum density required 
 for successful reheating. Assuming that the expansion rate being slow compared to the rate of evaporation (which is true) for the black holes density we must demand 
 \begin{equation}\label{constr3}
 T_{\rm reh}>T_{\rm BBN} \Rightarrow  \rho_{BH}(t^-_{ev}) > \frac{\pi^2\, g_{\rm reh}}{30}T_{\rm BBN}^4 \, ,
 \end{equation}
 where $g_{\rm reh}$ is the number of massless degrees
 of freedom for the reheated plasma. This black hole density  $\rho_{BH}(t^-_{ev})$ affects through the cosmic evolution (solving the system of differential equations of BD gravity) the initial black hole density $\rho_{BH}(t_{i})$.

Regarding the free parameter $N$, it is possible to calculate bounds too. 
The number of black holes must be at least $N_{min}$. this lower bound can be determined requesting 
\begin{equation}\label{constr4}
\rho_{BH}(t^-_{ev})=N\,m_{BH}\,H^3(t_{ev}) > \frac{\pi^2\, g_{\rm reh}}{30}T_{\rm reh}^4  
\end{equation}
setting the minimum allowed BH mass $m_{BHmin}$ (the lower bound from previous second constraint) we find the more strict lower bound for $N$
\begin{equation}\label{constr4b}
N > \frac{\pi^2\, g_{\rm reh}}{30\,m_{BHmin}}T_{\rm reh}^4 \, H^{-3}(t_{ev}) =N_{min}
\end{equation}

Another lower and upper bound for the number $N$, of black holes or the black hole density, $\rho_{BH}(t^-_{ev})$ i.e. the density before the sudden rapid evaporation, could be obtained demanding this density after the cosmic expansion dilution to be between the current cosmic critical density and much smaller value of the present cosmic radiation density, since in the most general case there may be physical process that dark energy interact with matter density and radiations density. However, since we "believe" and follow a BD conventional gravity only at high energies we need to know, for low energies, the modified BD gravity with varying $\omega$ in order to evaluate a range of values. 

Regarding, the Eddington luminosity. Setting the BH accretion luminosity equal to the Eddington limit gives us the maximum rate at which a black hole can accrete gas.  It is not known in detail what happens when we `feed’ a black hole with gas at a rate larger than the Eddington limit, but most probably, part of the gained mass will be ejected.  This limit is applicable to large astrophysical black holes with rotating accretion discs with opacity and viscosity. In our case PBHs are very small and very hot and around them we have a symmetric phase of massless particles. There is no Thomson scattering.   

All the presented parameters in the examples of the previous sections have values within the limits of the current section. For completeness, it worths to note that it is allowed of course to start with somewhat smaller than the lower bound black hole mass the accretion era as long as the accretion provides rapidly a mass within the allowed range.
 
%%%%%%%%%%%%%%%%%%%%%%%%%%%%%%%%%%%%%%%%%%%%%%%
\section{Primordial black holes mass spectrum}
%%%%%%%%%%%%%%%%%%%%%%%%%%%%%%%%%%%%%%%%%%%%%%%

In the previous sections we worked with the assumption that all the black holes have the same mass. Thus, it was possible to have some analytical solutions, to check if the model produces the observed baryon number and to set bounds on black holes' mass. Yet, it is more natural to assume that there is a spectrum of the initial masses. So, we are going to examine how this affects our model.

The two limits set in the previous section are still valid in the case of mass spectrum, since they refer to each one black hole's mass. PBHs with mass greater than the upper bound are not hot enough to thermalize their neighbourhood. If they have, on the other hand, mass less than the lower bound, then their lifetime is not long enough to form the domain wall where the baryogenesis would take place. Only the part of PBHs mass spectrum in the range between the two bounds contributes to the baryon number generation.

Eqs. (\ref{BRate}), (\ref{BR2}) for the baryon number created by a single PBH are still valid, but the total baryon asymmetry created by all PBHs is
\begin{equation}\label{Bspectrum}
b = \int ^{\infty} _{0} B\, N(m,t)\,dm\,,
\end{equation}
where $N$ is the number density of PBHs with masses from $m$ to $m+dm$. As a general conclusion, it suffices to state that the very efficient baryogenesis due to accretion remains unaffected from the presence of mass spectrum. Based on a certain cosmological scenario of the creation of PBHs one can estimate the exact baryon asymmetry straightforwardly. More details will follow concerning the relation of the black hole mass spectrum and the time evolution of the scale factor and the cosmic densities.

Next, we derive the equations governing the evolution of the spectrum of PBHs. We assume that the initial number density of the black hole spectrum is described by a power-law form, as in \cite{Carr75} and \cite{Barrow}. Thus, the initial number density of the PBHs with masses between $m_0$ and $m_0+dm_0$ is
\begin{equation}\label{initialnumber}
N(m_0)dm_0=A \,m_{0}^{-n}\,\Theta(m_0-m_c)\,dm_0 \, ,
\end{equation}
where $m_0=m(t=0)$ is the initial PBH mass. For the analytic calculations not to become unnecessarily complicated, we accept that all PBHs form at the same initial time $t_0$. We use $\Theta$ to introduce a cut-off mass $m_c$. This protects from divergences at the low-masses limit. Thus, we set $\Theta=1$ for $m>m_c$ and $\Theta=0$ for $m\leq m_c$. We assume that $m_c$ is proportional to the Planck mass, $m_c=k\,m_{pl}$, where the constant $k$ is arbitrary and has no dimensions. For the total energy density not to diverge at large masses, it has to be $n>2$. According to Carr \cite{Carr75}, initial density perturbations that produce PBHs in standard cosmology, indicate that $n$ is between 2 and 3. $A$ is the amplitude of the spectrum. Its units are such that $N(m_0)dm_0$ is number density.

The total number density of the black holes, as a function of time, is
\begin{equation}\label{initnd}
N(t)=\int ^{\infty} _{0} N(m,t)\,dm\,,
\end{equation}
and their total energy
\begin{equation}\label{energydensity}
\varrho_{BH}(t)=\int ^{\infty} _{0} N(m,t)\,m\,dm\,.
\end{equation}
Treating analytically the evolution of the mass spectrum considering both accretion and evaporation, was not possible. Yet, in our model the epoch when accretion is dominant is succeeded very quickly by an epoch when evaporation prevails, resulting in a reheated, radiation dominated universe. Thus, one can treat the two epochs separately.

%%%%%%%%%%%%%%%%%%%%%%%%%%%%%%%%%%%%%%%%%%%%
\subsection{Dominant accretion time period}

Here we will analyze the time period after primordial black hole creation where the accretion is important. Our aim is to calculate the evolution of black holes and radiation densities and the scale factor.

The factors that determine the PBHs mass spectrum evolution are not only the universe's expansion, but accretion also. The rate of gain, because of accretion, for a single black hole is given by Eq.(\ref{acc}). Solving it we get
\begin{equation}\label{massbhac}
m_{0}=\frac{1}{ m(t)^{-1}+16\,\pi\,f\,I_\rho} \, ,
\end{equation}
where $I_\rho=\int ^{t} _{0} G^2\,\rho_{rad}$\,.

Differentiating Eq.(\ref{massbhac}) with respect to $m_0$, we can have an expression for the evolution of the number density of PBHs with masses from $m$ to $m+dm$ at time $t$, combining it with Eq. (\ref{initialnumber}) (special care must be given for the jacobian factor). So, the evolution of the mass spectrum with time is
\begin{equation}\label{ndac}
N(m,t)dm=N(m_0,t)dm_0=A\, \big(\frac{1}{ m(t)^{-1}+16\,\pi\,f\,I_\rho}\big)^{2-n}\,\,m^{-2}\Theta(m-m_{ca}(t))\,dm\,,
\end{equation}
where $m_{ca}$ is the cut-off mass that is evolved from $m_c$:
\begin{equation}\label{mca}
m_{ca}(t)= \Big( \frac{1}{k \, m_{pl}} -16\,\pi\,f\,I_\rho \Big)^{-1}\,.
\end{equation}
One can see that, contrary to the evaporation epoch, the cut-off mass never becomes $0$.

The energy density rate is determined using the identity
\begin{equation}\label{identity}
\frac{d}{dx}\int ^{f(x)} _{g(x)} h(x,y)\,dy=\int ^{f(x)} _{g(x)}\frac{\partial h(x,y)}{\partial x} \,dy + h(x,f(x))
\frac{df(x)}{dx}-h(x,g(x))\frac{dg(x)}{dx}\,,
\end{equation}
The energy density of the radiation that is eaten by the PBHs and so is added to the black hole density $\varrho_{bh}$ (not the comoving), from time $t$ to $t+dt$, is $dE=\varrho_{bh}(t+dt)-\varrho_{bh}(t)=\frac{\partial \varrho_{bh}}{\partial t}\,dt$. The energy density rate, then, is calculated using also Eq. (\ref{identity}):
\begin{eqnarray}\label{dedtaccr}
\frac{dE}{dt}=\frac{d}{dt}\int_{0}^{\infty}N(m,t)\,m\,dm&=&\int_{m_{ca}}^{\infty}A(n-2)\big( m^{-1}+\xi \,I_\rho \big)^{n-3}\xi\,\,\frac{dI_\rho}{dt}  \,m^{-1}\,dm  \\ \nonumber
&& - A\,\big( m_{ca} ^{-1}+\xi \,I_\rho \big)^{n-2} \, m_{ca}^{-1}\,\frac{dm_{ca}}{dt}\,\Theta(m-m_{ca}(t)).
\end{eqnarray}
where $\xi=16\pi\,f$ and
\begin{equation}\label{dmca}
\frac{d\,m_{ca}}{dt}= \Big( \frac{1}{k \, m_{pl}} -\xi\,I_\rho \Big)^{-2}\,\xi\,\frac{dI_\rho}{dt} .
\end{equation}
The first term of Eq. (\ref{dedtaccr}) is actually the evolution of the spectrum. The second term is present because of the mass cut-off evolution. In the accretion era this term does not vanish, since evaporation is insignificant, compared to accretion. Then, we can have the full equations that determine the expansion, where the densities must be multiplied by $\alpha^{-3}$, in order to become comoving. 

The resulting set of equations is
\begin{equation}\label{com-b}
\rho_{BH}=\frac{1}{\alpha^3}\int ^{\infty} _{0} N(m,t)\,m\,dm \,,\,\,\,\rho=\rho_{BH}+\rho_{rad}
\end{equation}
\begin{equation}
\frac{dE_{com}}{dt}=\frac{1}{a^3}\frac{dE}{dt}\\
\label{expansionaccr-b}
\end{equation}
\begin{equation}
\dot{\rho}_{rad}=-4\frac{\dot{\alpha}}{\alpha}\rho_{rad}-|\frac{dE_{com}}{dt}|\,,
\label{expansionaccr2-b}
\end{equation}
since the kinetic pressure by the black holes is not important.

This set is supplemented by Eqs. (\ref{Fried}) and either (\ref{G}) for the first case or (\ref{G2}) for the second case. They are an integro-differential system, which is solved only numerically for various ranges of the parameters.

%%%%%%%%%%%%%%%%%%%%%%%%%%%%%%%%%%%%%%%%%%%%%%%%%%
\subsection{Dominant evaporation time period}

At some point in time accretion becomes less important than evaporation. This happens due to the ongoing expansion of the universe and, mainly, because the whole of the universe's radiation ends inside the PBHs, as we explained in the previous sections. From that time on, evaporation dominates the evolution of the black hole mass. The significance of this analysis lies in finding the modifications to the expansion rate, allowing the emergence of the conventional radiation expansion law. We aim to determine the deviations of the PBHs and of radiation densities and the evolution of the scale factor.

The evolution of the PBHs mass spectrum depends on the expansion of the universe and, more importantly, on the evaporation of the PBHs. The rate of mass loss of a single black hole, because of evaporation, is given by Eq.(\ref{ev}). Integrating it we get Eq.(\ref{evap}). Note that now the initial value $m_0=m_{max}$ is the maximum value of the black hole mass after the end of the dominant accretion time period.
\begin{equation}
m^3=m_0^3-\frac{3\,a_H}{256\pi^3}\,I_g
\end{equation}
where $I_g=\int_{0}^{t} G^{-2}dt$.
Then, we solve with respect to $m_0$ and differentiate. Thus,  we can have the evolution of the
black holes number density from $m$ to $m+dm$ at time $t$ from Eq.(\ref{initialnumber}). The evolved mass spectrum is given by
\begin{equation}\label{nd}
N(m,t)dm=A\, \left( m^3+ \frac{3\,a_H}{256\pi^3}\,I_g \right)^{-(n+2)/3}\,m^2 \,\Theta(m-m_{cr}(t))\,dm\,,
\end{equation}
where the cut-off mass is evolved, too:
\begin{equation}\label{mcr}
m_{cr}(t)= [(k \, m_{pl})^3 - \frac{3\,a_H}{256\pi^3}\,I_g ]^{1/3}\,.
\end{equation}
We can see that there is a time $t_{lim}$ that the cut-off mass becomes $0$.

The energy density that is emitted by the black hole as radiation from $t$ to $t+dt$
is estimated from Eq. (\ref{energydensity}). It is
\begin{equation}
dE=\varrho_{bh}(t)-\varrho_{bh}(t+dt)=-\frac{\partial \varrho_{bh}}{\partial t}\,dt\,.
\label{dE}
\end{equation}
where the quantities are not comoving. We can have the energy density rate using the identity Eq.(\ref{identity}). So, we find
\begin{eqnarray}
-\frac{dE}{dt}&=&\frac{d}{dt}\,\int ^{\infty} _{0} N(m,t)\,m\,dm \\\nonumber
&=&A\,\frac{\,a_H}{256\pi^3}\,(-n-2)\,\int ^{\infty} _{m_{c,max}} m^{3}\,\left(  m^3+ \frac{3\,a_H}{256\pi^3}\,I_g \right)^{-(n+5)/3}\, \frac{dI_g}{dt} \,dm \nonumber\\
&&-  A\,m_{cr}^3 \,\left(  m_{cr}^3+ \frac{3\,a_H}{256\pi^3}\,I_g \right)^{-(n+2)/3}\,\Theta(m-m_{cr}(t))\,\frac{dm_{cr}}{dt},
\label{dedt}
\end{eqnarray}
where
\begin{equation}
\frac{dm_{cr}}{dt}= \frac{-\,a_H}{256\pi^3}\,[(k \, m_{pl})^3 - \frac{3\,a_H}{256\pi^3}\,I_g ]^{-2/3}\,\frac{dI_g}{dt}
\end{equation}
and
\begin{equation}\label{max}
m_{c,max}(t)=max[0,m_{cr}(t)]\,.
\end{equation}
The first term in Eq. (\ref{dedt}) expresses the evolution of the spectrum and is the only non-zero term at late times. The second term of Eq. (\ref{dedt}) is present because of the time evolution of the mass cut-off. It is apparent that for times larger than $t_{lim}$ the lightest black holes completely evaporate and the $\Theta$ function causes this term to vanish.

In all the quantities calculated so far, the dilution from the expansion will have to be added; the comoving density is $\rho_{BH}=\varrho_{bh}\,a^{-3}$, and the comoving energy is $E_{com}=Ea^{-3}$.

Finally, the set of equations is the following
\begin{equation}\label{com}
\rho_{BH}=\frac{1}{a^3}\int ^{\infty} _{0} N(m,t)\,m\,dm \,,\,\,\,\rho=\rho_{BH}+\rho_{rad}
\end{equation}
\begin{equation}
\frac{dE_{com}}{dt}=\frac{1}{a^3}\frac{dE}{dt}\\
\label{expansionaccr}
\end{equation}
\begin{equation}
\dot{\rho}_{rad}=-4\frac{\dot{a}}{a}\rho_{rad}+|\frac{dE_{com}}{dt}|\,,
\label{expansionaccr2}
\end{equation}
since black holes exert unimportant kinetic pressure.

Eqs. (\ref{com}), (\ref{expansionaccr}), (\ref{expansionaccr2}), (\ref{Fried}) and either (\ref{G})  for the first case or (\ref{G2}) for the second case are an integro-differential system. Like in the dominant accretion time period, the equations system can be solved only numerically for various ranges of the parameters.

%%%%%%%%%%%%%%%%%%%%%%%%%%%%
\section{Discussion and Conclusions}
%%%%%%%%%%%%%%%%%%%%%%%%%%%%

A very efficient baryogenesis mechanism was proposed in the early cosmic evolution of a Universe with Brans-Dicke gravity. According to the studied scenario very small primordial black holes born at the end of the BD - field domination era ($\sim 10^{-35}sec$) create the observed baryon number.

For the case that the coupling constant $\omega$ is constant and equal to $10^4$ (which is the lowest possible value for the present time) we have found that primordial black holes with initial mass $m_i \ge 10^{27} GeV$, accrete radiation from their surroundings intensively, leading to almost complete PBH domination, even if PBHs density was initially only 1/100,000 of the universe density. However, the maximum of PBH mass should not exceed $\sim 10^{32} GeV$.

For greater values of $\omega$, which is closer to General Relativity, accretion is less intense: for $\omega=10^{10}$ it is $m_i \ge 10^{28} GeV$. For lower values of $\omega$ accretion is enhanced. The initial PBH mass can be as low as $10^{17}GeV$ for $\omega=1$, increased then by a factor of $10^{15}$ to lead to BH domination.

The final produced baryon to entropy asymmetry depends on the black holes number N and the CP-violation angle $\Delta \theta_{CP}$. For $\omega=10^{4}$ and $m_i=10^{27}GeV$, for example, it is $N\Delta \theta_{CP}\geq2.4\times10^{33} $ for the observed $b/s \geq 6\times 10^{-10}$. Thus for reasonable values of $N$, $\Delta \theta_{CP}$ can be within the limits of phenomenologically accepted two Higgs doublet models.

We proved that BD gravity, due to enhanced accretion, can naturally provide black holes domination in the early Universe and at the same time, efficient baryogenesis for smaller CP violating angles compared to the case of the conventional gravity of General Relativity.

The proposed baryogenesis happens for a very short period in the very early Universe during for which we worked under the simplified assumption that $\omega$ remains more-less constant. During the cosmic evolution, in general $\omega$ can change value, i.e. it can be larger in order to meet the observational constraints. There are various Brans-Dicke type models, \cite{BDW}, with a varying omega exhibiting an additional contribution to the change of the gravitational constant over time due to $\omega$ evolution. Of course $\omega$ should take a correct large enough phenomenological value from the nucleo-synthesis era and afterwards. 
In all the baryogenesis successful scenaria that we have found like for example in the case with $\omega$ varying from $1$ at early times to $10^4$ today, there are model dependent constraints in this time evolution of $\omega$. 
These constraints however are dependent on the specific generalised Brans-Dicke or scalar tensor gravity model. 
Naturally, any constraint on $G$ or its derivative derived assuming FRW background or simple BD cosmology degrades in the context of generalised BD or scalar tensor theory (as the scalar field will also source the background dynamics, thereby influencing the expansion rate, and at the same time being responsible for the time variation of $G$). Some self-consistent analyses for a simple BD gravity, find $\omega >300$ from BBN alone \cite{BDBBN}. So this is a value of $\omega$ we must have at BBN era while for today the most strong constraint, as we have mentioned, is $\omega >10^4$ (Shapiro time delay measurements by the Cassini satellite).
Many generalized BD models or scalar tensor models can both satisfy the observational constraints and give a big variation of $\omega$ like the one we need in our scenario. 
As an example we mention the work \cite{Majumdar:2007kx} which studies a specific generalised BD model that is compatible with the observational constraints and the $\omega$ can even start from an hypothetical initial $\omega=10^{-14}$ and end to a today value $\omega=10^4$. 

As a future work it would be interesting to adopt a specific generalized BD model to study the present proposed baryogenesis mechanism and find the constraints that would apply in the free parameters of the model.  

It is also worth studying the ideas presented in this work for Asymptotic Safe Gravity \cite{AS}, since it shares some similar properties to Brans-Dicke models. Another interesting question is to analyse how initial anisotropic or inhomogeneous backgrounds (with small anisotropies/inhomogeneities that smooth out later) affect the mechanism \cite{anisotr}.

\newpage

\section{Acknowledgments}
We acknowledge enlightening discussions with A. S. Majumdar and B. Nayak.
V. Zarikas acknowledges the support of Orau Grant SOE2019010, No. 110119FD4534, “Quantum gravity at astrophysical scales.”

\end{document}